\documentclass[12pt, reqno]{article}
\usepackage{natbib}
\usepackage{color}
\usepackage{graphicx}
\usepackage{bm}
\usepackage{amsmath,amssymb}
\usepackage{accents}
\usepackage{color}
\usepackage{subfigure}
\newcommand{\me}{e}
\newcommand{\mH}{H}
\newcommand{\mi}{i}

\newcommand{\dif}{d}

\renewcommand{\vec}[1]{\bm{#1}}

\def\XXint#1#2#3{{\setbox0=\hbox{$#1{#2#3}{\int}$}
     \vcenter{\hbox{$#2#3$}}\kern-.5\wd0}}

\newcommand\hatphi{\skew3\hat{\phi}}      

\newcommand{\doublehat}[1]{%
    \hat{\hat{#1}} }

\newcommand{\new}[1]{{\color{black}{#1}}}

\title{On wavenumber spectra for sound within subsonic jets}

\author{%
  A. Agarwal%
\thanks{Lecturer, Department of Engineering, University of Cambridge (aa406@cam.ac.uk)}
  \quad S. Sinayoko%
\thanks{Brunel Research Fellow, ISVR, University of Southampton (s.sinayoko@soton.ac.uk)}
  \quad R. D. Sandberg%
\thanks{Professor, AFM, University of Southampton (r.d.sandberg@soton.ac.uk)}
}

\date{\today}

\graphicspath{{img/}}

\begin{document}
\maketitle
\begin{abstract}
This paper clarifies the nature of sound spectra within subsonic jets.
Three problems, of increasing complexity, are presented.
Firstly, a point source is placed in a two-dimensional plug flow and the sound field is obtained analytically.
Secondly, a point source is embedded in a diverging axisymmetric jet and the sound field is obtained by solving the linearised Euler equations.
Finally, an analysis of the acoustic waves propagating through a turbulent jet obtained by direct numerical simulation is presented.
In each problem, the pressure or density field are analysed in the frequency-wavenumber domain.
It is found that acoustic waves can be classified into three main frequency-dependent groups.
A physical justification is provided for this classification.
The main conclusion is that, at low Strouhal numbers, acoustic waves satisfy the d'Alembertian dispersion relation.
\end{abstract}

\section{Introduction}
Our initial motivation for understanding the sound spectra in jets came from
the article by \cite{Goldstein:2005vt} in which he proposed that it
may be possible to identify the ``true'' sources of noise in jets if
the radiating and non-radiating components could be separated. It is
possible to achieve this separation for the Euler equations linearised
about either a steady uniform base flow \citep{CHU:1958to} or a steady
parallel flow \citep{Agarwal:2004wt}. Unfortunately, the separation
techniques presented in these papers cannot be applied to full
nonlinear Navier-Stokes equations and hence are not useful for
realistic jets.

\cite{Sinayoko:2011ij} showed that filtering in the
frequency-wavenumber domain is an effective technique for separating
radiating and non-radiating components in subsonic jets. Their
filtering technique relied on the dispersion relation
$k=|\omega|/c_{\infty}$ (where $k$ denotes the magnitude of the
wavenumber, $\omega$ the angular frequency and $c_{\infty}$ the
farfield speed of sound) satisfied by acoustic waves radiating to a
quiescent farfield. But inside the jet we can have waves that travel
supersonically relative to the ambient medium. In this paper, we
define acoustic waves in jets as those satisfying the dispersion
relation $k_{z} \leq |\omega|/c_{\infty}$, where $k_{z}$ denotes the
axial wavenumber. In other words, in the axial direction, acoustic
waves travel either upstream ($k_z \leq 0$) or downstream ($0 \leq k_z
\leq |\omega|/c_{\infty}$; in the downstream case, the axial phase
speed is therefore sonic or supersonic. The characterization of
acoustic waves by supersonic axial phase speed was used
by~\cite{Freund:2001vw},~\cite{Cabana:2008ht},~\cite{Tinney_Jordan_2008}
and~\cite{obrist_directivity_2009}. The results presented in this
paper support this definition of acoustic waves.

\begin{figure}
	\begin{center}
		\includegraphics{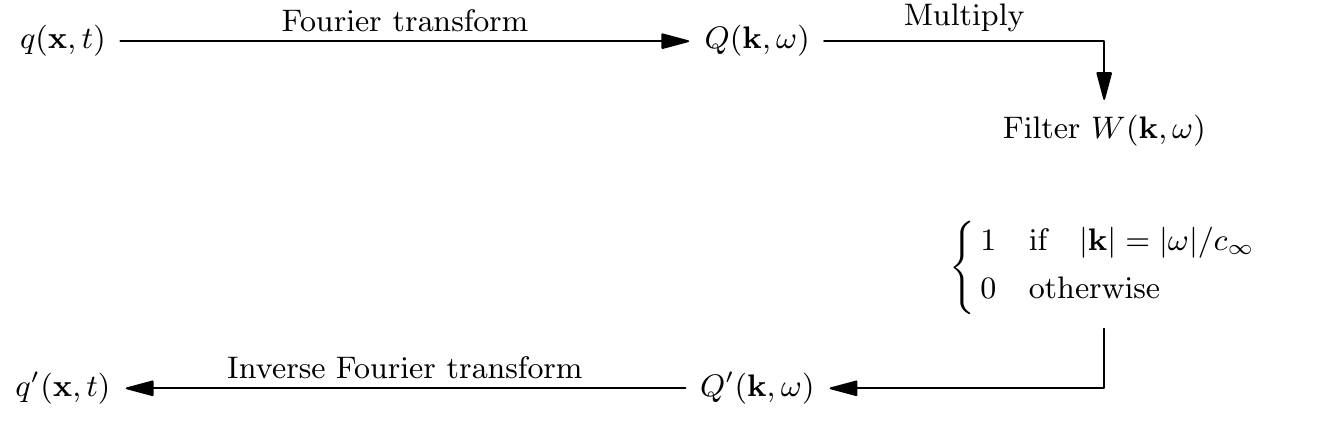}
	\end{center}
\caption{Algorithm for filtering out the radiating field. For numerical
    implementation, $W(\vec{k}, \omega)$ has a finite width (see
    \cite{Sinayoko:2011ij} for details)}
\label{fig:algo}
\end{figure}

\begin{figure}
  \centering
  \begin{tabular}{cc}
  \subfigure[$\rho_{\omega=1.2}( \mathbf x, t_0)$]{\includegraphics{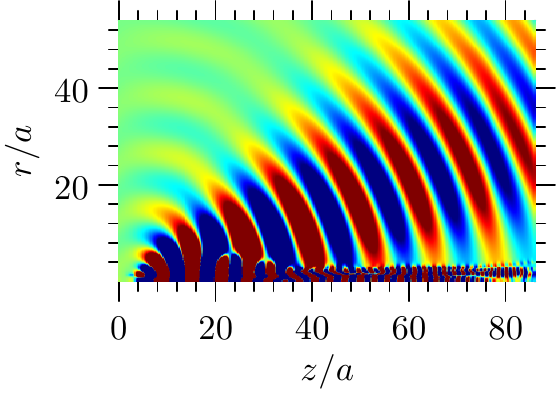}} &
  \subfigure[$P( \mathbf k, 1.2)$]{\includegraphics{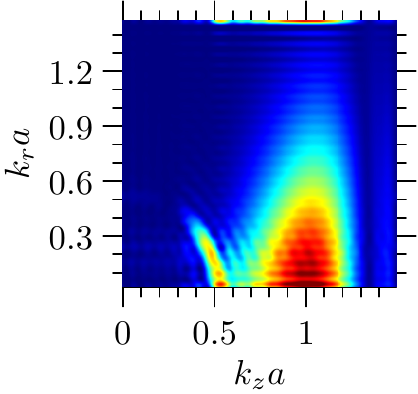}}  \\
  \subfigure[$\rho'_{\omega=1.2}( \mathbf x, t_0)$]{\includegraphics{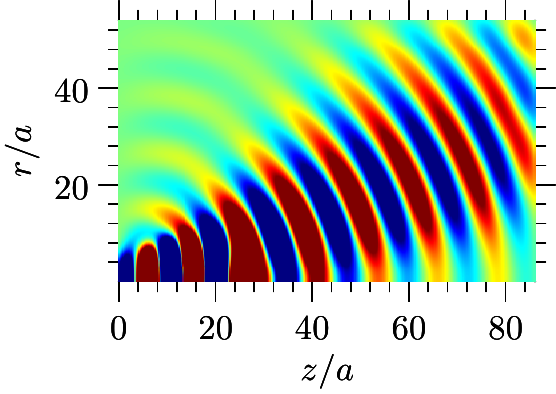}} &
  \subfigure[$P'( \mathbf k, 1.2)$]{\includegraphics{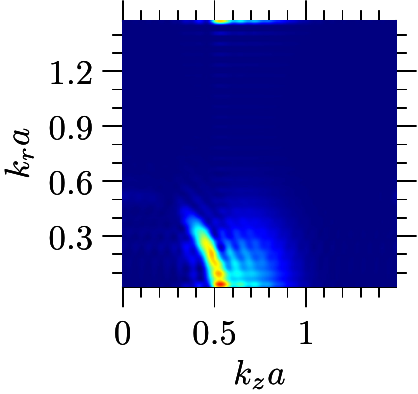}}  \\
  \subfigure[$\overline{\rho}_{\omega=1.2}( \mathbf x, t_0)$]{\includegraphics{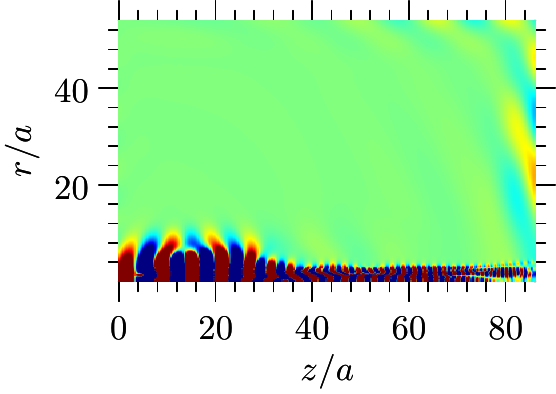}} &
  \subfigure[$\overline{P}( \mathbf k, 1.2)$]{\includegraphics{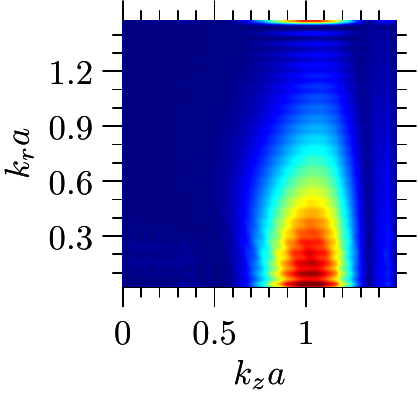}}
\end{tabular}
  \caption{Density field in a turbulent jet of exit radius $a$ plotted in the
  physical domain
  (snapshot in left column) and wavenumber domain (magnitude of Fourier
  transform in right column), with colour range $[-5\times 10^{-5},
  5\times 10^{-5}]$ and
  $[0, 0.5]$ respectively, for a given normalized frequency $\omega=1.2$ and
  time $t_0$ (c.f. equation~\eqref{eq:notation}).  The top row shows the density field $\rho$, the middle row the
  radiating field $\rho'$ and the bottom row the non-radiating field
  $\overline{\rho}$.}
  \label{fig:laminar}
\end{figure}

The filtering technique is represented diagrammatically in figure
\ref{fig:algo}. The radiating part $q'(\vec{x},t)$ of a flowfield
variable $q(\vec{x},t)$ can be obtained by convolving $q$ with an
appropriate filter function $w(\vec{x},t)$, which is defined in the
frequency-wavenumber domain ($W(\vec{k},\omega)$).
\cite{Sinayoko:2011ij} considered a model problem in which the base
flow corresponding to the experiment of the Mach 0.9, Re 3600 jet by
\cite{Stromberg:1980ur} was excited by two instability waves at
nondimensional frequencies of 2.2 and 3.4. These waves interact
nonlinearly to produce acoustic waves at the difference frequency of
1.2. The density field at frequency 1.2 is shown in figure
\ref{fig:laminar} (a). In this problem, both acoustic and hydrodynamic
waves are being generated. The Fourier transform of this field, $P(\vec{k},
\omega)$, is shown in figure \ref{fig:laminar} (b). Multiplying
$P(\vec{k},\omega)$ with $W(\vec{k},\omega)$ as defined in figure
\ref{fig:algo} gives $P'(\vec{k},\omega)$, which is shown in figure
\ref{fig:laminar} (d). The Fourier transform of the remaining field,
$\bar{P}(\vec{k},\omega) = P - P'$, is shown in figure
\ref{fig:laminar} (f). The corresponding density fields in the
space-time domain are obtained by applying the inverse Fourier
transforms and are shown in figures \ref{fig:laminar} (c) and (e).
Radiating components have captured all the acoustic waves. Clearly the
acoustic waves have been separated from the hydrodynamic waves.
However, the efficacy of the filter is puzzling as it is based on the
dispersion relation for sound propagation in a uniform quiescent
medium. Inside the jet we do not have a quiescent medium, so how can
this dispersion relation separate acoustic waves both outside and
inside the jet?

In order to answer this question, we have constructed  a simple model problem
for sound radiation from a point source in a two-dimensional plug flow (\S
\ref{sec:model}).  We show that, for this problem, it is possible to obtain an
analytical expression for the Fourier transform for both the axial and
cross-stream directions. This is a crucial step in obtaining the spectral
characteristics of sound propagation and it enables us to understand and
explain the observed acoustic wavenumber spectra for different frequencies.
The solution to this problem also indicates how to identify acoustic waves for
more general (turbulent) jets.  In \S \ref{sec:diverging} we consider a more
general problem of sound radiation from a point source in a diverging
cylindrical jet and in \S \ref{sec:turbulent} we identify the acoustic waves
using data obtained from a DNS of a Mach 0.84, Re 7200 turbulent jet.

Even though our motivation for identifying acoustic waves in turbulent jets
stems from a particular application as mentioned above, this work can be used
in other ways. For example, the flow filtering technique defined here could be
used to separate convecting and propagating components in a jet. This can help
define various source models or correlate the nearfield hydrodynamic data to
the acoustic farfield to identify the noise producing regions in the jet. The
technique can also be used to correctly identify the radiating part of
Lighthill's source term (\citet{Freund:2001vw}, \citet{Cabana:2008ht},
\citet{Sinayoko:2012bt}).

\section{Model problem} \label{sec:model}

\begin{figure}
  \begin{center}
    \includegraphics[scale=0.8]{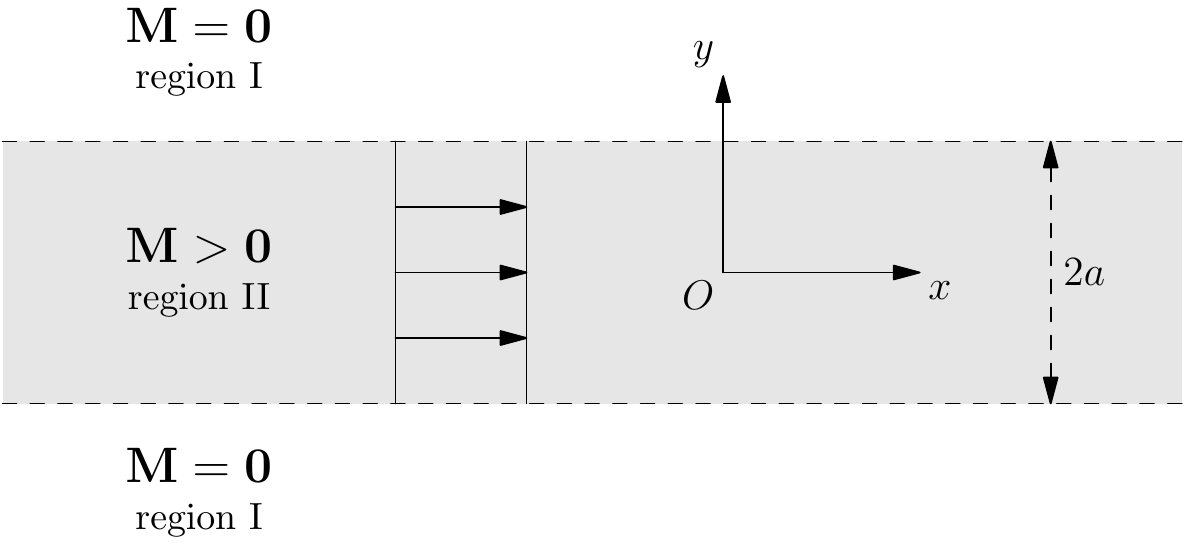}
  \end{center}
  \caption{Schematic sketch of \new{a} point source located at the origin inside a
  two-dimensional plug flow of width $2 a$. \new{The flow Mach number is $M>0$ for $|y| < a$ (region II) and $M=0$ otherwise (region I).}}
  \label{fig:schematic}
\end{figure}

Consider the problem of a time-harmonic monopole point mass source,
$\rho_o \delta(\vec{x}) \exp(-\mi \omega t)$ ($\rho_o$ denotes ambient
density), embedded in a plug flow.
Several authors (e.g. \cite{Morgan:1975te}, \cite{Mani:1972va})
have considered the problem of farfield sound radiation from a point source in
axisymmetric jets. The main difference between their analysis and ours is that
we seek the spectral content in the frequency-wavenumber domain instead of
the farfield characteristics of sound in the physical domain.

For simplicity, we consider a two-dimensional problem. The problem set up is
described in figure~\ref{fig:schematic}. Assuming an $\exp(-\mi \omega t)$
response ($\phi(\vec{x},t) = \phi(\vec{x};\omega) \exp(-\mi \omega t)$), the linear velocity potential $\phi_I$ for small disturbances
satisfies, in region I (outside the jet),
\begin{equation}
	\label{eqn:helm1}
	\nabla^2 \phi_I + \kappa^2 \phi_I = 0,
\end{equation}
and in region II, inside the jet,
\begin{equation}
	\label{eqn:helm2}
	\nabla^2 \phi_{II} - \left( -\mi \kappa + M \frac{\partial }{\partial x}
	\right)^2 \phi_{II} = \delta(x)\delta(y),
\end{equation}
where $\nabla^2$ denotes the Laplacian operator, $\kappa = \omega/c$ is the
acoustic wavenumber, and $c$ is the speed of sound, which is uniform for the
present problem. Because the velocity potential and pressure are symmetric
about the mid-plane axis of the jet ($y=0$), it is sufficient to solve the
problem for $y\ge0$. Continuity of pressure at $y=a$ requires that $p =
-\rho_o D \phi/ D t$ be continuous (D/Dt denotes material derivative),
therefore
\begin{equation}
	\left(-\mi \kappa + M \frac{\partial }{\partial x}\right) \phi_{II} (x, a) =
	-\mi \kappa \phi_I (x, a).
	\label{eqn:PresCont2}
\end{equation}
The kinematic constraint requires that particle displacement $\eta$ at the
interface be continuous. Therefore,
\begin{equation}
	\frac{\partial }{\partial y}\phi_I (x, a) =
		\frac{\partial }{\partial t}\eta(x,a),
	\label{eqn:eta1}
\end{equation}
\begin{equation}
	\frac{\partial  }{\partial y}\phi_{II}(x,a) = \frac{\partial }{\partial t}\eta(x,a) +
	M c \frac{\partial }{\partial x}\eta(x,a).
	\label{eqn:eta2}
\end{equation}

\begin{figure}
  \begin{center}
    \includegraphics{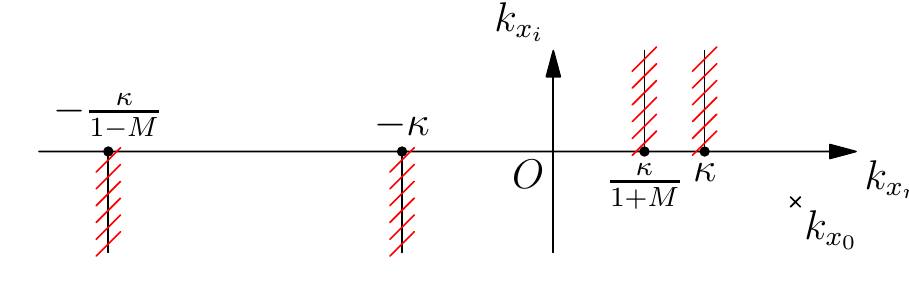}
  \end{center}
  \caption{Location of the branch cuts for $\beta$ and $\gamma$ (hatched
  lines) and the Kelvin-Helmholtz instability pole, $k_{x_0}$, in the complex
  $k_x$ domain.}
  \label{fig:kx_domain}
\end{figure}

Applying the Fourier transform in $x$\new{,} defined by
\begin{equation}
	\hatphi(k_x) = \int_{-\infty}^\infty \phi(x) \me^{-\mi k_x x} \dif x,
	\label{eqn:FT}
\end{equation}
to equations (\ref{eqn:helm1}) and (\ref{eqn:helm2}), we get,
\begin{equation}
	\frac{\dif^2 \hatphi_I}{\dif y^2} + (\kappa^2 - k_x^2) \hatphi_I = 0,
	\label{eqn:helm1a}
\end{equation}
\begin{equation}
	\frac{\dif^2 \hatphi_{II} }{\dif y^2} + \left((\kappa-k_x M)^2 - k_x^2\right)
	\hatphi_{II}  = \delta(y).
	\label{eqn:helm2a}
\end{equation}
Application of the Fourier transform in the axial direction to
Eqs.~(\ref{eqn:PresCont2}) -- (\ref{eqn:eta2}) gives
\begin{equation}
	(\kappa-k_x M) \hatphi_{II} (k_x, a) = \kappa \hatphi_I(k_x, a),
	\label{eqn:PresCont3}
\end{equation}
\begin{equation}
	\kappa \frac{\dif \hatphi_{II} }{\dif y} (k_x, a) = (\kappa - k_x M)
	\frac{\dif \hatphi_I} {\dif y} (k_x,a).
	\label{eqn:eta3}
\end{equation}
Let $\beta^2 = \kappa^2 - k_x^2$ and $\gamma^2 = (\kappa-k_x M)^2 - k_x^2$. The
locations of the branch cuts for $\beta$ and $\gamma$ are shown in figure
\ref{fig:kx_domain}.
The branch of the square roots are chosen such that both $\beta$ and $\gamma$
are equal to $\kappa$ for $k_x=0$.  Acoustic waves propagate to the farfield only
when $\beta$ is real, i.e. when $|k_x| < \kappa$. Therefore,
we will focus on this range of wavenumbers.  In region I for outgoing
waves to infinity
\begin{equation}
	\hatphi_I = A \me^{\mi \beta y}.
	\label{eqn:hatphi1}
\end{equation}
Taking into account the symmetry about the mid-plane axis, the solution in region II is given by
\begin{equation}
	\hatphi_{II}  = 2 B \cos(\gamma y) - \frac{\mi}{2} \frac{\me^{\mi \gamma y}}{\gamma}.
	\label{eqn:hatphi2}
\end{equation}
Application of conditions (\ref{eqn:PresCont3}) and (\ref{eqn:eta3}) yields
\begin{eqnarray}
	A &=& -\frac{\mi \kappa \exp(-\mi a \beta) (\kappa-M k_x )}{2 \Delta(\kappa,k_x,\gamma)},\\
	B &=& \frac{\exp(\mi a \gamma ) \left(\kappa^2 \gamma -\beta
		     (\kappa-M k_x )^2\right)}{4 \gamma \mi \Delta(\kappa,k_x,\gamma)},
	\label{eqn:AB}
\end{eqnarray}
where $\Delta(\kappa, k_x, k_y) = \beta  \cos (a k_y ) (\kappa-M k_x )^2-\mi \kappa^2 k_y
\sin (a k_y )$.  Note that for $M=0$  we recover the free-field Green's function of the
Helmholtz equation.

Equation $\Delta(\kappa,k_x,\gamma) = 0$ is the dispersion relation for the
hydrodynamic wave. The roots of this equation represent poles of $\hat{\phi}$
in the complex $k_x$ domain. For the present problem there is one root,
$k_{x_0}$, which is located in the lower-half $k_x$-plane and is associated
with a Kelvin-Helmholtz instability wave; it is purely hydrodynamic
\citep{Agarwal:2004wt} and does not affect our analysis. If our problem had a
pipe (or two splitter plates in 2D) the Kelvin-Helmholtz instability wave
would have an amplitude given by the edge condition, usually an unsteady Kutta
condition (see for example, \cite{Cri85}). Again the Kelvin-Helmholtz wave
would be subsonic and hence would not interfere with our analysis.

Defining the Fourier transform in $y$ by
\begin{equation}
	\doublehat{\phi}(k_x,k_y) = \int_{-\infty}^\infty \hatphi(k_x, y) \me^{-\mi k_y
	y} \dif y,
	\label{eqn:FTy}
\end{equation}
the Fourier transform of the pressure field can be written as
\begin{eqnarray}
	\doublehat{p}(k_x, k_y) = -2 \mi \rho_o \int_0^\infty \left[ \kappa
	\hatphi_{II} (k_x, y) \mH(a-y) + \right.\nonumber\\
     \left. (\kappa - k_x M) \hatphi_I(k_x, y) \mH(y-a) \right]
	\cos (k_y y) \,\dif y.
  \label{FTpy}
\end{eqnarray}
This integral can be evaluated analytically:
\begin{eqnarray}
	\doublehat{p}(k_x, k_y) &=& \frac{\rho_o(\kappa-M k_x )}{\Delta (\kappa, k_x, \gamma)}
	\left[ \frac{\mi\left( \Delta(\kappa, k_x, k_y) - \Delta(\kappa, k_x, \gamma)\right)}
	{k_y^2-\gamma^2} + \right.  \nonumber\\
	& & \left. \frac{\kappa^2 \left(k_y \sin (a k_y) + i \beta
	\cos (a k_y)\right)}{k_y^2 - \beta^2} \right].
    \label{eqn:FTp}
\end{eqnarray}

\subsection{Frequency-wavenumber spectra}

The acoustic-wave solution in the physical domain can be obtained by
applying the inverse Fourier transforms to Eq.~(\ref{eqn:FTp})
(for details on the geometry of the Fourier integration contours $F_x$ and
$F_y$ and the implications on causality, see \cite{Agarwal:2004wt})
\begin{equation}
	p(x, y; \omega) =
	\int_{F_x} \frac{d k_x}{2 \pi} e^{i k_x x}
	\int_{F_y} \frac{d k_y}{2 \pi} e^{i k_y y}
	\doublehat{p}(k_x, k_y).  \label{eqn:invFL}
\end{equation}
If we look at
the $k_y$ integral, $\doublehat{p}(k_x, k_y)$, from Eq.~(\ref{eqn:FTp}), has
two terms.  The second term has poles at $k_y = \pm \beta$. Using the method
of residues, it can be shown that only these poles contribute to the integral.
Therefore, regardless of the frequency, only the wavenumbers that satisfy the
dispersion relation
is $k_x^2 + k_y^2 = \kappa^2$ contribute to the integral. We refer to this as the radiation circle.  The
other term in the integrand has zeroes in the denominator at $k_y = \pm
\gamma$, which \new{corresponds to an} ellipse in the wavenumber domain.  Note
that these zeroes do not represent poles as the numerator also goes to
zero at $k_y = \pm \gamma$.  Therefore, the contribution from the
integrand is more complicated for this term. Further insight can be obtained
by plotting the integrand as a function of frequency.  Recall that
$\kappa=\omega/c$ and from hereon, for brevity, the reduced frequency
$\kappa a$ is referred to as frequency.

\begin{figure}
	\centerline{\includegraphics{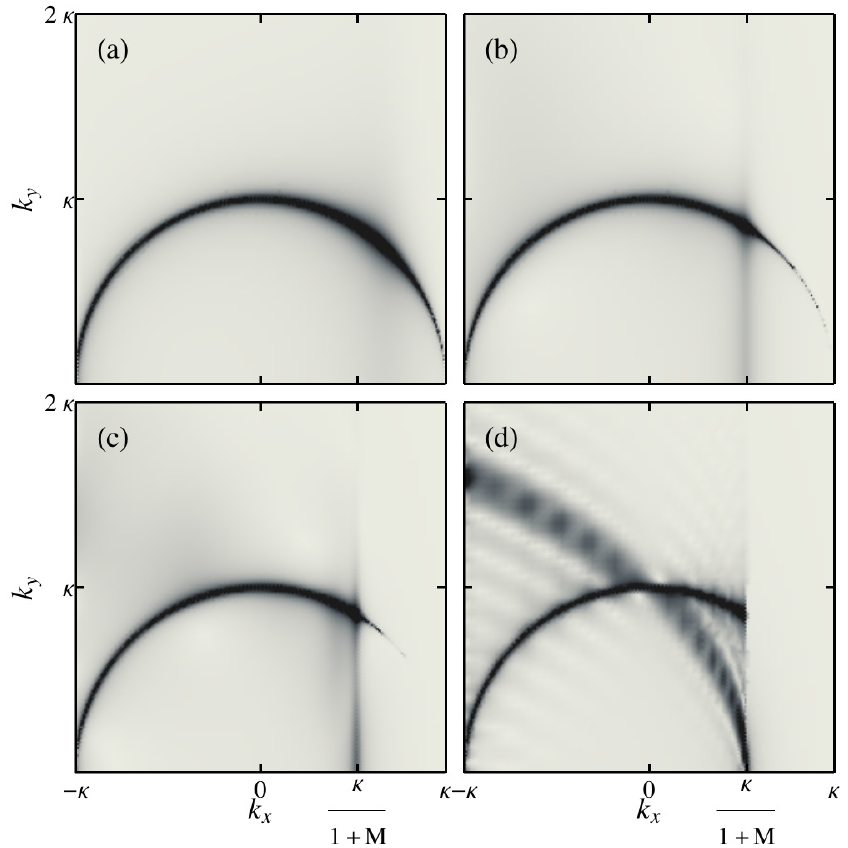}}
	\caption{Wavenumber spectra of the pressure field, $|\doublehat{p}|$, from
	a point source in a jet (set-up of figure \ref{fig:schematic}) at frequency,
	(a) $\kappa a = 0.1 \pi$, (b) $\kappa a = \pi$, (c) $\kappa a = 2 \pi$, (d) $\kappa a = 10
	\pi$. The density plot uses a linear scale from 0 (white) to $10 \pi /\kappa$
	(black).}
	\label{fig:spectra}
\end{figure}

Figure \ref{fig:spectra} shows the wavenumber spectra $|\doublehat{p}(k_x,
k_y)|$ for $M = 0.9$ for four different frequencies. At low frequencies \new{($\kappa a \ll 1$)} most
of the energy is concentrated around the radiation circle \new{(figure~\ref{fig:spectra}(a))}.  For higher
frequencies (\new{$ \kappa a = O(1)$}), the energy is concentrated along the radiation circle as well,
but there is a small amount of energy around the vertical line $k_x =
\kappa/(1+M)$ \new{(figures~\ref{fig:spectra}(b) and~\ref{fig:spectra}(c))}. For very high frequencies \new{($\kappa a \gg 1$)} we see the radiation circle and a
part of the ellipse $k_y^2 = \gamma^2$ \new{(figure~\ref{fig:spectra}(d))}. We observe some ringing around the
ellipse.

For jet noise another useful non-dimensional frequency is the Strouhal number,
$St$, based on the jet diameter and exit velocity. It can be shown that $St =
\kappa a / (\pi M)$. High-speed jet noise peaks at $St \sim 0.2$ ($\kappa a =
0.2 M \pi$). This suggests that for filtering out acoustic waves in a jet,
around the peak radiation frequency, one need not worry about the ellipse in
figure \ref{fig:spectra} (d).  For sound radiation near the peak frequency, the
dispersion characteristics are very similar to that of the ordinary wave
equation. This explains why \cite{Sinayoko:2011ij} obtained a
good separation of acoustic and hydrodynamic fields by using a filter based on
the dispersion characteristics of the ordinary wave equation.

\subsubsection*{Low frequencies}

A mathematical justification for this low-frequency result can be obtained as
follows. Assume $\kappa a \ll 1$.
For acoustic waves, the wavenumbers
$k_x$ and $k_y$ that satisfy the dispersion relation are of the same
order of magnitude as $\kappa$, so that $\gamma a \ll 1$ and
$k_y a \ll 1$. In Eq.~(\ref{eqn:FTp}) if we expand the trigonometric functions
in a power series up to the second order, it simplifies to
\begin{eqnarray*}
	\doublehat{p}(k_x, k_y) \approx \frac{\rho_o
	\beta  (\kappa-M k_x ) \left[\kappa^2 (\mi - a \beta ) + O\left( (a \kappa)^4 \right)
	\right]}{ \Delta(\kappa, k_x, \gamma) (k_y^2-\beta^2 ) }
	\label{eqn:FTpsmall}
\end{eqnarray*}
From this equation it is clear that at low
frequencies, the dispersion relation for the problem is $k_y^2 - \beta^2 = 0$,
i.e. $k_x^2 + k_y^2 = (\omega/c)^2$, which is the dispersion relation for
sound propagation through a uniform medium at rest. This indicates that
mean flow has a negligible effect on sound propagation at low frequencies.
This has been observed experimentally by \cite{CJCG12}.

The results can be interpreted by considering the potential field
$\hat{\phi}(k_x, y)$ given by Eqs.~(\ref{eqn:hatphi1}) and
(\ref{eqn:hatphi2}).  Figure \ref{fig:comparison}(a) shows the associated
pressure field $\hat{p}(k_x, y)$ as a function of $y$ for $k_x = \kappa/(1+M)$
for two different values of $\kappa a$, $0.1 \pi$ and $\pi$.

For $k_x = \kappa/(1+M)$, $\gamma=0$ and from Eq.~(\ref{eqn:hatphi2}) it is
clear that $\hat{p}$ does not have a wave-like solution for $y \leq a$. For
brevity, we consider only the real part of $\hat{p}$, which can be shown to be
a constant with respect to $y$ in the limit $\gamma \to 0$.  For $y>a$, the
response is given by the harmonic function $\exp(\mi \beta y)$.  At low
frequencies (e.g., $\kappa a=0.1 \pi$), the constant part (when $y < a$) is a
very small part of a wavelength (figure \ref{fig:comparison}(a)) and therefore,
when we Fourier transform this field, most of the energy is concentrated
around $k_y = \pm \beta$.  A similar argument applies to other values of $k_x$.
Therefore, the presence of the jet has a negligible effect on the wavenumber
spectrum at low frequencies.

\subsubsection*{Mid-range frequencies}

As the frequency increases, the region in the $y$-domain over which the
pressure field is a constant (when $k_x = \kappa/(1+M)$) occupies a larger
part of the wavelength (see the dashed line in figure \ref{fig:comparison} (a)).
This has a significant impact on the Fourier transform (in $y$) of $\hat{p}$.
Instead of being concentrated just around $k_y = \pm \beta$, the energy in the
$k_y$ space gets distributed over a range of $k_y$ values. The same reasoning
applies to other values of $k_x$ close to $\kappa/(1+M)$. \new{This explains the vertical patch in the
spectra around $k_x = \kappa/(1+M)$ in figures \ref{fig:spectra} (b)
and (c).}

Away from these values, in the range $0 < k_x < \kappa/(1+M)$ , unless
the frequency is very high, the pressure field inside the jet is very similar
to that outside the jet. \new{This is illustrated in figure \ref{fig:comparison}(b), which shows
the pressure field $\hat{p}(k_x, y)$ for $k_x = 0.5 \kappa/(1+M)$, for $\kappa a=0.1\pi$ (solid line) and $\kappa a=\pi$ (dashed line)}. Therefore, for $0 < k_x < \kappa/(1+M)$, the
Fourier transform of the pressure field is concentrated around the radiation
circle $k_y = \pm \beta$.

The energy in the $k_y$ direction appears to be contained mainly inside the
radiation circle around $k_x = \kappa/(1+M)$. This can be explained as
follows.  The difference between the constant field for $y<a$ and the sinusoid
$\exp(\mi \beta y)$, that would exist in the absence of a jet, is
\new{related} to the Heaviside function $H(a-y)$. The Fourier transform of
this function is given by the sinc function, $\mathrm{sinc}(k_y a)$. Most of
the energy of this function is contained in the first lobe $k_y a < \pi$.
This is why there is little energy outside the radiation circle for $\kappa a
= \pi$ (figure \ref{fig:spectra}(b)). Thus the energy content inside the
radiation circle is a consequence of the pressure field inside the jet. Also
the phase speed of the content inside the radiation circle is supersonic
relative to a laboratory reference frame. This energy content inside the
radiation circle represents modes trapped within the jet.  \new{The
implication for flow decomposition into radiating and non-radiating components
is that some acoustic components lie within the radiation circle as
well($\kappa^2 < (\omega / c)^2$).}

From Eq.~(\ref{eqn:hatphi2}) and the definition of $\gamma$, it can be seen
that for $k_x > \kappa/(1+M)$, we get an exponentially decaying response
inside the jet. Physically \new{this} represents a subsonically propagating wave
inside the jet that leads to an evanescent wave. This explains why the spectrum
decays for $k_x > \kappa/(1+M)$ for moderate to high frequencies. The decay
can also be explained using ray theory, which predicts a shadow region in a
wedge in the forward direction with half angle $\cos^{-1}[1/(1+M)]$
(\cite{Morse:1968}). A point on the radiation circle determines the direction of
sound radiation to the far field. The angle to the jet axis is given by
$\cos^{-1}(k_x/\kappa)$ (\cite{Goldstein:2005vt}). Thus the shadow region
predicted by ray theory is in agreement with the region of decay on the
radiation circle in Figs.  \ref{fig:comparison}(b) -- (d).

\subsubsection*{High frequencies}

\begin{figure}
	\centerline{\includegraphics{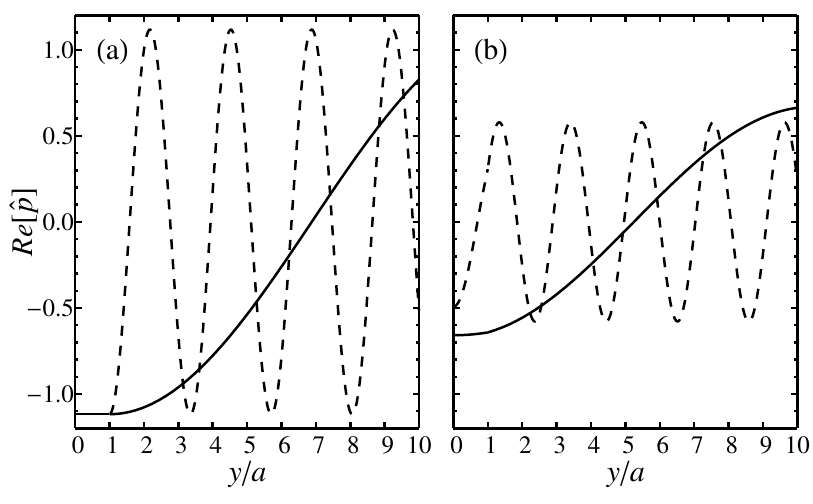}}
	\caption{Comparison of the pressure field $\hat{p}(k_x, y)$ at $\kappa a = 0.1 \pi$
	(solid line) and $\kappa a = \pi$ (dashed line) for  (a) $k_x = \kappa/(1+M)$,
	(b) $k_x = 0.5 \kappa/(1+M)$.}
	\label{fig:comparison}
\end{figure}

For very high frequencies, we see a combination of the radiation circle and an
ellipse.  At high frequencies we would get several wavelengths inside the jet
($y<a$, compare with figure \ref{fig:comparison}(b)). If $a$ is large (say,
infinite), then the spectra would correspond to that of a point source in a
uniformly moving medium, which is an ellipse. This is what we see in figure
\ref{fig:spectra}(d).  The ringing is because of the finiteness of $a$, which
in the wavenumber domain,
results in the convolution of the sinc function with the radiation ellipse.

\section{Diverging jet}	\label{sec:diverging}
In order to consider the effect of three-dimensionality and divergence of a jet on the results presented in the preceding section, we seek the acoustic field radiating from a monopole source embedded in diverging axisymmetric mean flow at $(z/a, r/a) = (3, 0)$.
The mean flow was obtained from the DNS of a Mach 0.84 and Re 7200 turbulent jet embedded in a co-flow of Mach 0.2. The full details of the DNS are available in \cite{Sandberg11e}, and its sound field is analysed in the next section.
The acoustic field was obtained by solving the linearised Euler equations.

\begin{figure}
  \centering
  \begin{tabular}{cc}
  \subfigure[$\rho_{ka=0.1\pi}( \mathbf x, t_0)$]{\includegraphics[height=4cm]{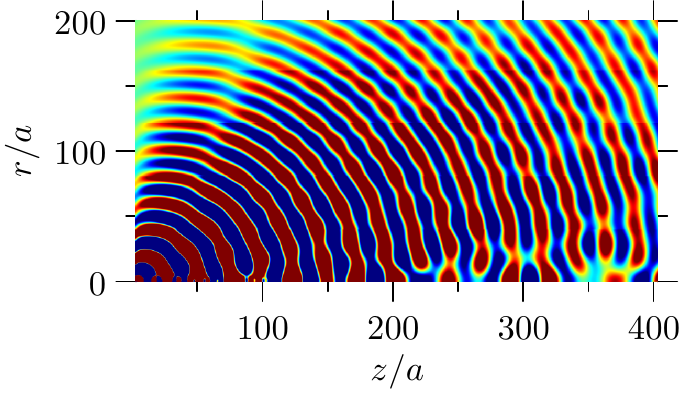}} &
  \subfigure[$P( \mathbf k, 0.1\pi)$]{\includegraphics[height=4cm]{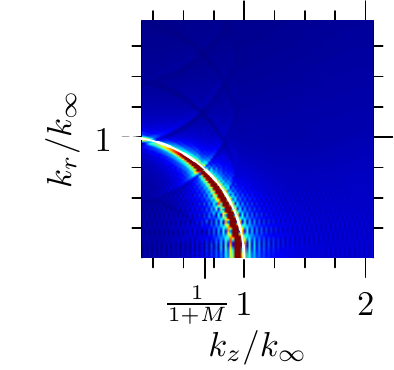}} \\
  \subfigure[$\rho_{ka=\pi}( \mathbf x, t_0)$]{\includegraphics[height=4cm]{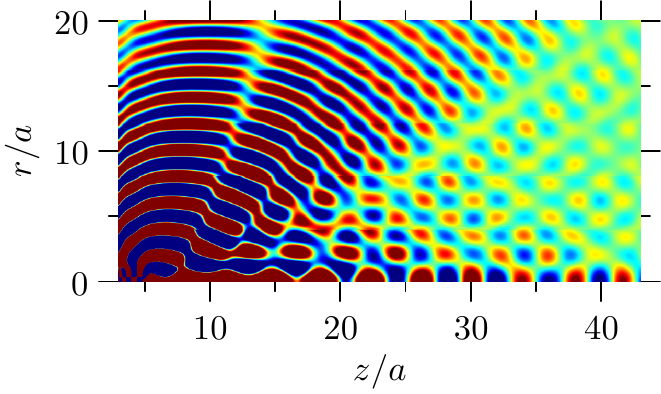}} &
  \subfigure[$P( \mathbf k, \pi)$]{\includegraphics[height=4cm]{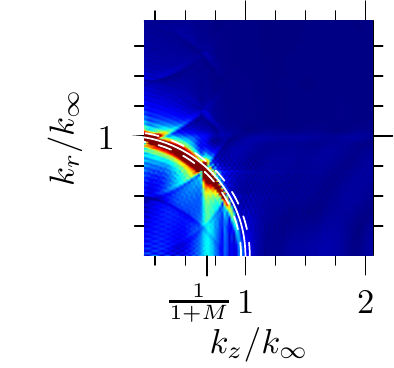}} \\
  \subfigure[$\rho_{ka=2\pi}( \mathbf x, t_0)$]{\includegraphics[height=4cm]{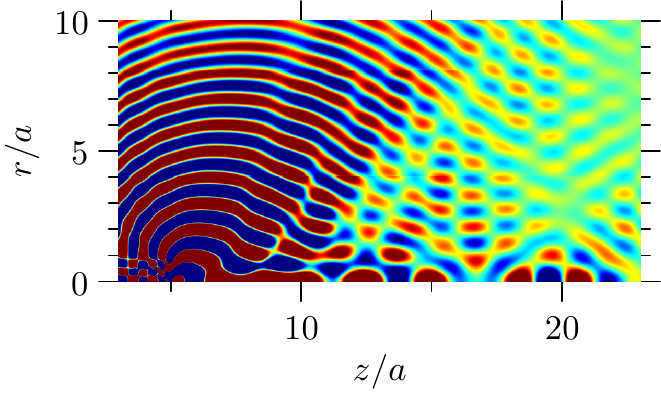}} &
  \subfigure[$P( \mathbf k, 2\pi)$]{\includegraphics[height=4cm]{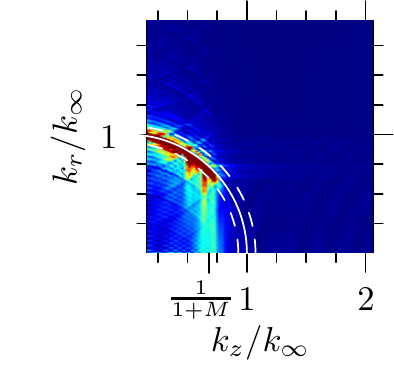}}\\
\end{tabular}
\caption{Density field radiating from a monopole in a diverging  mean flow of a turbulent jet plotted in the physical domain (snapshot in left column) and wavenumber domain (magnitude of Fourier transform in right column), with colour range $[-5.10^{-5}, 5. 10^{-5}]$ and $[0, 0.15]$ respectively, for normalized frequencies $ka/\pi=0.1, 1$ and $2$ and an arbitrary time $t_0$ (c.f. equation~\eqref{eq:notation}).  In the right column, the arc shows the radiation circle $|\mathbf k|=\omega/c_\infty$. The dashed arcs are at $|\mathbf k|=\omega/c_\infty \pm \Delta  k/2$, where $\Delta  k = \sqrt{\Delta k_z^2 + \Delta k_r^2}$ is the grid step in the wavenumber domain.}
\label{fig:diverging}
\end{figure}

\begin{figure}
  \centering
  \includegraphics[]{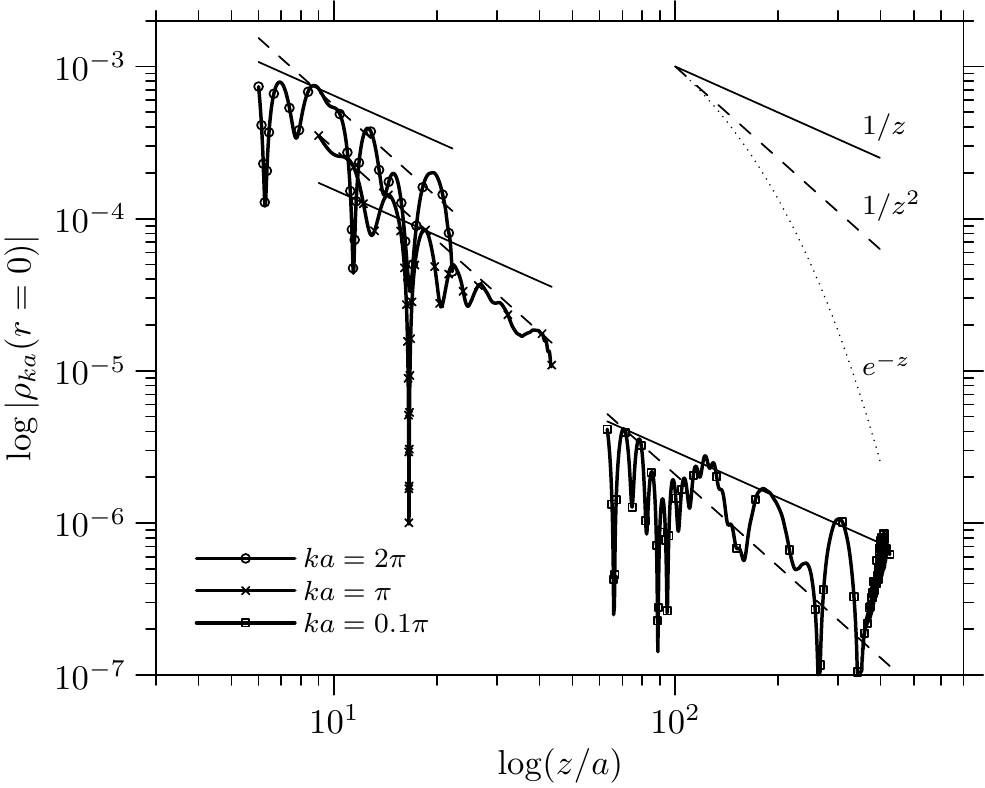}
  \caption{Profiles of the density field and the associated decay rate along $r=0$ for the data of figures~\ref{fig:diverging}(a, c, e).}
  \label{fig:decay_rates_diverging_jet}
\end{figure}

Figures \ref{fig:diverging} (a), (c) and (e) show the density field at frequencies of $k a = 0.1 \pi$, $\pi$ and $2 \pi$, respectively.
The notation $\rho_{ka}(\mathbf x, t_0)$ describes a linear combination of the real and imaginary parts of the temporal Fourier transform of $\rho$ at frequency $ka$, defined as
\begin{equation}
  \label{eq:notation}
\rho_{ka} (\mathbf x, t_0) =  \frac{1}{\pi} \left(\Gamma_r(\mathbf x, ka) \cos(\omega t_0) - \Gamma_i(\mathbf x, ka) \sin(\omega t_0)\right),
\end{equation}
where $\Gamma = \Gamma_r + i \Gamma_i$ is the temporal Fourier transform of $\rho(\mathbf x, t)$. This allows to follow the evolution of the density field at frequency $ka$ with time.
Figures \ref{fig:diverging} (b), (d) and (f) show the corresponding wavenumber spectra.
Comparing these figures for the case of monopole in the plug flow at the same
frequencies (figures \ref{fig:spectra}(a), (b) and (c)), we can see that the
diverging jet
has not changed the nature of the spectra. They look very similar. We see the
radiation circle, the shadow region and the vertical line around $k_x =
\kappa/(1+M)$, which can again be identified as trapped waves. These waves are
clearly visible around the centerline in the physical domain (figure
\ref{fig:diverging} (c) and (e)). The trapped modes propagate to the farfield,
as can be seen by looking at the density along the axis of the jet (figure
\ref{fig:decay_rates_diverging_jet}). The decay rate is a polynomial of order
less than 2.5 for all frequencies and not exponential and thus, these waves are
not evanescent.

The presence of trapped modes inside the radiation circle implies that the
filter $|\mathbf k|=|\omega|/c_\infty$ (figure~\ref{fig:algo}) does not
capture all the acoustic waves within the jet. To do so, one may use the
condition $|\mathbf k| \leq |\omega|/c_\infty$ instead.

\section{Turbulent jet} \label{sec:turbulent}
\begin{figure}
  \centering
  \begin{tabular}{cc}
  \subfigure[$\rho_{St=1.1}( \mathbf x, t_0)$]{\includegraphics{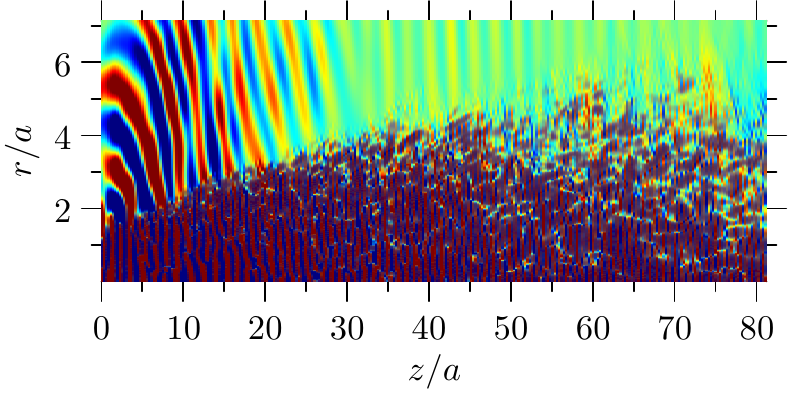}} &
  \subfigure[$P( \mathbf k, 1.1)$]{\includegraphics{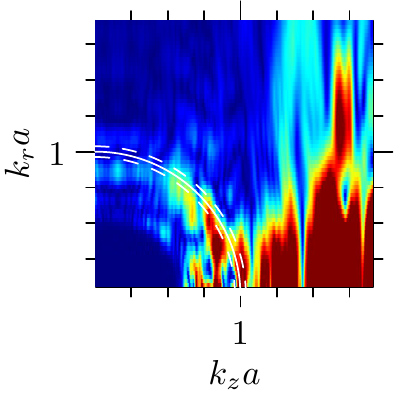}}  \\
  \subfigure[$\rho'_{St=1.1}( \mathbf x, t_0)$]{\includegraphics{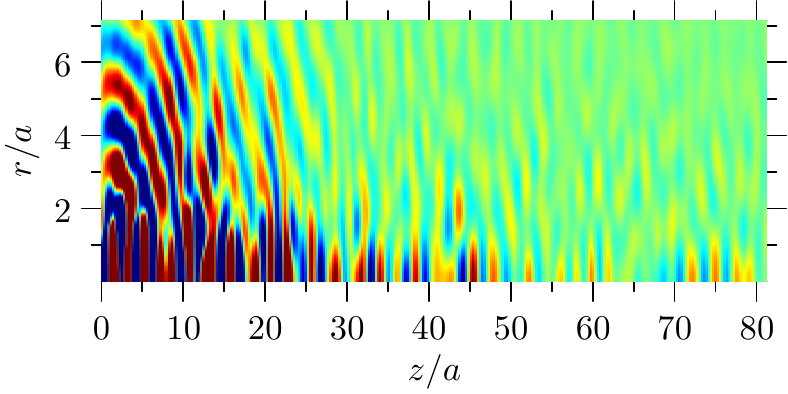}} &
  \subfigure[$P'( \mathbf k, 1.1)$]{\includegraphics{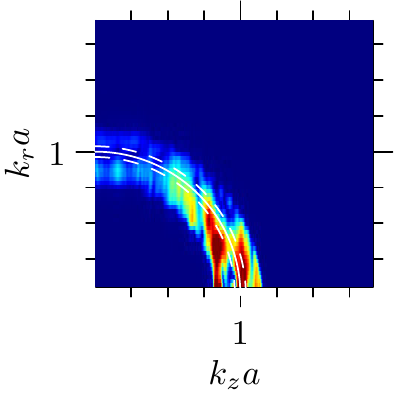}}  \\
  \subfigure[$\overline{\rho}_{St=1.1}( \mathbf x, t_0)$]{\includegraphics{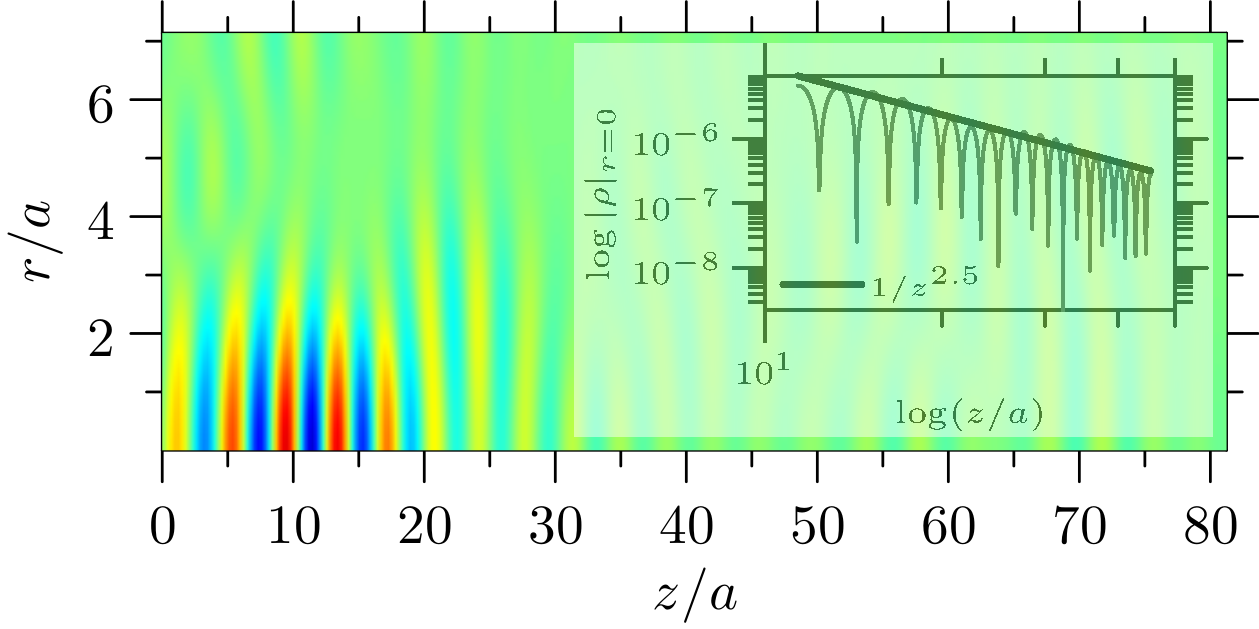}} &
  \subfigure[$\overline{P}( \mathbf k, 1.1)$]{\includegraphics{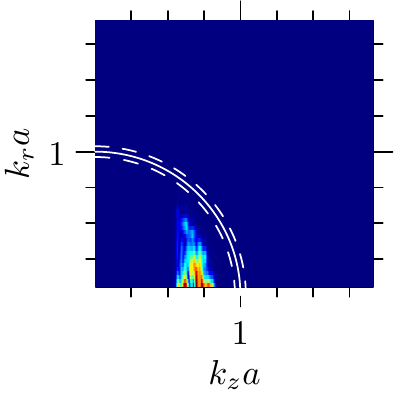}}
\end{tabular}
\caption{Density field in a turbulent jet plotted in the physical domain (snapshot in left column) and wavenumber domain (magnitude of Fourier transform in right column), with colour range $[-8\times 10^{-6}, 8\times 10^{-6}]$ and $[0, 0.5]$ respectively, for a given normalized frequency $St=1.1$ and time $t_0$ (c.f. equation~\eqref{eq:notation}).  The top row shows the density field $\rho$, the middle row the radiating field $\rho'$ and the bottom row the non-radiating field $\overline{\rho}$. In the right column, the arc shows the radiation circle $|\mathbf k|=\omega/c_\infty$. The dashed arcs are at $|\mathbf k|=\omega/c_\infty \pm \Delta  k/2$, where $\Delta  k = \sqrt{\Delta k_z^2 + \Delta k_r^2}$ is the grid size in the wavenumber domain.}
\label{fig:turbulent}
\end{figure}

The density field from the DNS~\cite{Sandberg11e} of a Mach 0.84 and Re 7200 turbulent jet embedded in a co-flow of Mach 0.2 is shown in figure \ref{fig:turbulent} (a)
at a Strouhal number (based on the jet exit diameter and velocity) of 1.1. This corresponds to $\kappa a = 0.92 \pi$.

The usual Fast Fourier Transform algorithm is inconvenient for computing the
temporal Fourier transform, since it requires storing the complete time
history of the three dimensional density field. Here, Goertzel's algorithm
(\citeyear{goertzel1958algorithm}) was used instead, as it consumes the time
history one frame at a time. No special windowing, i.e. a rectangular window,
was used to keep the main lobe as narrow as possible. However, a small amount
of spectral leakage from low frequencies was observed, resulting in highly
supersonic components ($|\mathbf k| \leq .6 k_\infty)$ appearing in the
wavenumber plots. These supersonic components are un-physical and have been
filtered out by means of a low pass filter of the form
\begin{equation}
  \label{eq:low_pass_filter}
  W_L(\mathbf k) = \frac{1}{2} \left(1 + \tanh((|\mathbf k| - k_0) / (\sigma k_\infty) \right),
\end{equation}
where $k_\infty = \omega / c_\infty$, $k_0 / k_\infty = 0.6$, $\sigma = 0.1$.

Figure \ref{fig:turbulent} (b) shows the Fourier transform of the density
field. If we filter only along the radiation circle as for the laminar jet,
the density field in the Fourier domain and the corresponding field in the
physical domain are shown in figures \ref{fig:turbulent} (d) and
\ref{fig:turbulent} (c) respectively. A qualitative comparison between figures
\ref{fig:turbulent} (a) and \ref{fig:turbulent} (c) shows that this filter
based on the radiation circle captures most of the acoustic waves. But it can
be seen from figure \ref{fig:turbulent} (b) that there is a significant amount
of energy inside the radiation circle. From the above analysis we expect that
this region of the spectrum should contribute to trapped waves along the axis
of the jet. Figure \ref{fig:turbulent} (f) shows the filtered spectra when
only the interior of the radiation circle is considered. The corresponding
density field in the physical domain is shown in figure \ref{fig:turbulent}
(e). As expected, the waves propagate close to the axis of the jet.
These trapped waves propagate to the far field,
as can be seen in the inset of figure \ref{fig:turbulent} (e), which
shows the magnitude of the density field along the centerline of the jet. As
in the preceding section, we get a polynomial decay rate for these trapped
waves.

Based on the above discussions, we can predict that if we are interested in
the far field acoustic wave radiating at a particular angle $\theta$, then we
can obtain this by filtering on the radiation circle in the
frequency-wavenumber domain for the same polar angle in the $k_r - k_z$ plane.
This is not surprising and was shown theoretically by \cite{Goldstein:2005vt}.
However for mid to high frequencies, this procedure would miss the effect of
trapped waves that propagate close to the axis of the jet.

It is worth noting that the wavenumber-frequency make-up of the
acoustic spectra is different than that of the turbulence. The
turbulent kinetic energy spectrum is generally obtained from numerical
data by the application of the three-dimensional Fourier transform in
space. The resulting spectrum contains a broad range of wavenumbers
and tends to be maximum for low wavenumbers. Since the radiation
circle is also located over low wavenumbers, one may think that there
is no separation between the turbulence and acoustic spectra. This is
not the case because the radiation circle is defined for a particular
frequency: a Fourier transform in time is required in addition
to the Fourier transform in space. In that case, for a given
frequency, only a narrow band of wavenumbers make up the turbulent
spectrum. In general, that narrow band would lie outside the radiation
circle (corresponding to subsonic propagation speeds). For example, to
compute the turbulence spectrum from experimental data, it is common
practise to measure at a single point in space and to Fourier
transform the signal in time. The Fourier transform in space is then
obtained by invoking Taylor's hypothesis: assuming a frozen pattern of
turbulence convected at a local flow speed $U$, the wavenumber and
frequency are related by $\omega/k = U$. Thus, for a given frequency,
we are picking out a single wavenumber of the turbulence spectrum that
corresponds to this dispersion relation. For a subsonic jet this
corresponds to a subsonic wave that lies outside the radiation circle.
Thus, for the example problem considered here the turbulence spectrum
corresponds to the energy content to the right of the radiation circle
in \ref{fig:turbulent} (b).

\section{Conclusions}
Most of the acoustic waves radiating from a jet satisfy the d'Alembertian dispersion relation $k = \omega/c_\infty$, i.e. they lie on the radiation circle in the frequency-wavenumber domain. This validates the radiation criterion proposed by~\cite{Goldstein:2005vt} and used by~\cite{Sinayoko:2011ij}.  

At low Strouhal numbers (e.g. $St \leq 0.5$), acoustic waves lie mainly on 
the radiation circle ($k = \omega/c_\infty$). This explains
why the dispersion relation based on an ordinary wave equation was sufficient
to filter out the acoustic waves even inside the jet, shown in figure
\ref{fig:laminar}(c).

At mid-Strouhal numbers ($St \sim 1$), some acoustic waves are trapped in the jet. These trapped waves can be classified as acoustic based on the observation that:
\begin{enumerate}
\item they propagate to the far field ;
\item they have supersonic phase speed $k < \omega/c_\infty$ (in a subsonic jet the hydrodynamic waves and the energy associated with turbulent structures convect at subsonic speeds).
\end{enumerate} 
These trapped acoustic waves can be identified by using the criterion $k \leq \omega/c_\infty$. Alternatively, one can use the axial wavenumber, $|k_x| \leq \omega/c_\infty$ since there are usually no hydrodynamic components such that $|k_x| < \omega/c_\infty$ and $|k_r| > \omega/c_\infty$. These would represent (unphysical) waves with axial wavelengths larger than acoustic waves travelling at subsonic speeds at high angles to the downstream jet axis. An advantage of this approach is that it does not require computing the radial Fourier transform.

Finally, at high Strouhal numbers ($St \gg 1$), some acoustic waves lie around
the radiation ellipse corresponding to the dispersion relation for waves
propagating through a flow of Mach number equal to the average convection Mach
number. These waves can therefore extend outside the radiation circle.

Although the results presented in this paper are for high Mach number subsonic 
jets, the solution of the plug flow problem at low Mach numbers indicates that 
the conclusions are valid for low Mach number flows as well. 

The above conclusions were shown to hold for sound propagation through a
time-averaged diverging jet and a turbulent jet. Similar results could likely
be obtained for other turbulent flows, such as mixing layers and wakes. If the
flow field in the far field is non-quiescent, which would be the case for a
mixing layer, then the radiation circle turns into a radiation ellipse. 



\begin{thebibliography}{}

\bibitem[Agarwal et~al., 2004]{Agarwal:2004wt}
Agarwal, A., Morris, P., and Mani, R. (2004).
\newblock {Calculation of sound propagation in nonuniform flows: suppression of
  instability waves}.
\newblock {\em AIAA J.}, 42(1):80--88.

\bibitem[Cabana et~al., 2008]{Cabana:2008ht}
Cabana, M., Fortun{\'e}, V., and Jordan, P. (2008).
\newblock {Identifying the radiating core of Lighthill's source term}.
\newblock {\em Theoretical and Computational Fluid Dynamics}, 22(2):87--106.

\bibitem[Cavalieri et~al., 2012]{CJCG12}
Cavalieri, A.~V., Jordan, P., Colonius, T., and Gervais, Y. (2012).
\newblock Axisymmetric superdirectivity in subsonic jets.
\newblock {\em Journal of Fluid Mechanics}, 704:388--420.

\bibitem[Chu and Kovasznay, 1958]{CHU:1958to}
Chu, B. and Kovasznay, L. (1958).
\newblock {Non-Linear Interactions in a Viscous Heat-Conducting Compressible
  Gas}.
\newblock {\em J. Fluid Mech.}, 3(5):494--514.

\bibitem[Crighton, 1985]{Cri85}
Crighton, D.~G. (1985).
\newblock The kutta condition in unsteady flow.
\newblock {\em Annual Review of Fluid Mechanics}, 17(1):411--445.

\bibitem[Freund, 2001]{Freund:2001vw}
Freund, J.~B. (2001).
\newblock {Noise sources in a low-Reynolds-number turbulent jet at Mach 0.9}.
\newblock {\em Journal of Fluid Mechanics}, 438:277--305.

\bibitem[Goertzel, 1958]{goertzel1958algorithm}
Goertzel, G. (1958).
\newblock {An algorithm for the evaluation of finite trigonometric series}.
\newblock {\em The American Mathematical Monthly}, 65(1):34--35.

\bibitem[Goldstein, 2005]{Goldstein:2005vt}
Goldstein, M. (2005).
\newblock {On identifying the true sources of aerodynamic sound}.
\newblock {\em J. Fluid Mech.}, 526:337--347.

\bibitem[Mani, 1972]{Mani:1972va}
Mani, R. (1972).
\newblock {A moving source problem relevant to jet noise}.
\newblock {\em J. Sound Vib.}, 25(2):337--347.

\bibitem[Morgan, 1975]{Morgan:1975te}
Morgan, J. (1975).
\newblock {The interaction of sound with a subsonic cylindrical vortex layer}.
\newblock {\em Proc. R. Soc. A}, 344(1638):341--362.

\bibitem[Morse and Ingard, 1968]{Morse:1968}
Morse, P. and Ingard, K. (1968).
\newblock {\em Theoretical acoustics}.
\newblock Princeton University Press.

\bibitem[Obrist, 2009]{obrist_directivity_2009}
Obrist, D. (2009).
\newblock Directivity of acoustic emissions from wave packets to the far field.
\newblock {\em Journal of Fluid Mechanics}, 640:165--186.

\bibitem[Sandberg et~al., 2012]{Sandberg11e}
Sandberg, R., Suponitsky, V., and Sandham, N. (2012).
\newblock {DNS of compressible pipe flow exiting into a coflow}.
\newblock {\em Int. J. Heat Fluid Fl.}, 35:33--44.

\bibitem[Sinayoko and Agarwal, 2012]{Sinayoko:2012bt}
Sinayoko, S. and Agarwal, A. (2012).
\newblock {The silent base flow and the sound sources in a laminar jet}.
\newblock {\em The Journal of the Acoustical Society of America}, 131(3):1959.

\bibitem[Sinayoko et~al., 2011]{Sinayoko:2011ij}
Sinayoko, S., Agarwal, A., and Hu, Z. (2011).
\newblock {Flow decomposition and aerodynamic sound generation}.
\newblock {\em J. Fluid Mech.}, 668:335--350.

\bibitem[Stromberg et~al., 1980]{Stromberg:1980ur}
Stromberg, J., McLaughlin, D., and Troutt, T. (1980).
\newblock {Flow field and acoustic properties of a Mach number 0· 9 jet at a
  low Reynolds number}.
\newblock {\em Journal of sound and vibration}, 72(2):159--176.

\bibitem[Tinney and Jordan, 2008]{Tinney_Jordan_2008}
Tinney, C.~E. and Jordan, P. (2008).
\newblock The near pressure field of co-axial subsonic jets.
\newblock {\em Journal of Fluid Mechanics}, 611:175–204.

\end{thebibliography}
\end{document}